\documentclass[final,5p,times,twocolumn]{elsarticle}
\usepackage{hyperref}
\usepackage{graphicx}
\usepackage{wrapfig}
\usepackage{caption}

\title{Study of MicroPattern Gaseous detectors with novel nanodiamond based photocathodes for single photon detection in EIC RICH}
\author[add-ts-infn]{J.~Agarwala\corref{ictp}}
\author[add-trieste]{C.~Chatterjee\corref{cor}}
\ead{chandradoy.chatterjee@ts.infn.it}
\author[add-ba-cnr]{G. Cicala}
\author[add-ts-infn]{A.~Cicuttin\corref{ictp}}
\author[add-trieste]{P.~Ciliberti}
\author[add-ts-infn]{M.L.~Crespo\corref{ictp}}
\author[add-ts-infn]{S.~Dalla~Torre}
\author[add-ts-infn]{S.~Dasgupta}
\author[add-ts-infn]{M.~Gregori}
\author[add-ts-infn]{S.~Levorato}
\author[add-ts-infn]{G.~Menon}
\author[add-ts-infn]{F.~Tessarotto}
\author[add-ba]{A.Valentini}
\author[add-ba-cnr]{L. Velardi}
\author[add-ts-infn]{Y.~Zhao}

\cortext[ictp]{also at Abdus Salam ICTP, Trieste,Italy}
\cortext[cor]{Corresponding author}

\address[add-ts-infn]{INFN Trieste, Trieste, Italy}
\address[add-trieste]{University of Trieste and INFN Trieste, Trieste, Italy}
\address[add-ba]{University Aldo Moro of Bari and INFN Bari, Bari, Italy}
\address[add-ba-cnr]{ P.Las.M.I. Lab @CNR-NANOTEC and INFN, Bari, Italy }

\date{September 2018}
\begin{document}
\begin{abstract}
 Identification of high momentum hadrons at the future EIC is crucial, gaseous RICH detectors are therefore viable option. Compact collider setups impose to construct RICHes with small radiator length, hence significantly limiting the number of detected photons. More photons can be detected in the far UV region, using a windowless RICH approach. 
\par
QE of CsI degrades under strong irradiation and air contamination. Nanodiamond based photocathodes (PCs) are being developed as an alternative to CsI. Recent development of layers of hydrogenated nanodiamond powders as an alternative photosensitive material and their performance, when coupled to the THick Gaseous Electron Multipliers (THGEM)-based detectors, are the objects of an ongoing R\&D. 
 We report about the initial phase of our studies.  
\end{abstract}
\maketitle 

\section{Introduction}
The proposal of an Electron Ion Collider(EIC)~\cite{EIC} is made to address many fundamental aspects of particle physics, in particular those related to the origin of
nucleon mass and spin, and the properties of dense gluon systems. Efficient Particle IDentification (PID) at high momenta, namely identification of hadrons above $6$-$8~GeV/c$ will play an important role for the proposed studies of Quantum ChromoDynamics (QCD) at such EIC. A gaseous Ring Imaging CHerenkov(RICH) detector is therefore an obvious choice. However, the number of Cherenkov photons generated in a light gaseous radiator is limited. Currently, good number of photons is achieved by using long radiators. The compact architecture of the experimental setup at the EIC collider is a constraint for this technique. According to the Frank-Tamm distribution, in the far ultraviolet (UV) domain ($\sim120$~nm), the density of generated Cherenkov photon is larger. This implies the detection of photons in the very far UV range, which is not accessible with standard fused-silica windows, since they are not transparent for wavelengths below 165~nm. $MgF_{2}$ or $CaF_{2}$ windows have cut-off at a much shorter wavelength, but to have larger number of detected photons a windowless RICH is an option. The windowless concept implies to the use of gaseous photon detectors operated with the radiator gas itself~\cite{windowless-RICH}, like the PHENIX HBD~\cite{HBD}.
\par 
MicroPattern Gaseous Detector (MPGD)-based Photon Detectors (PD) have recently been demonstrated as effective devices for single photon detection in Cherenkov imaging counters thanks to the application for the upgrade of COMPASS RICH-1~\cite{THGEM-RICH2018}. These detectors have a hybrid architecture, where two layers of THGEMs are followed by a Micromegas stage; the first THGEM also acts as PC substrate 
and is coated with a CsI film.
\par 
CsI is nowadays an obvious choice to convert photons in the far UV domain. Albeit efficiently detecting far UV photons, CsI has two concerning limitations. Firstly, delicate handling is required: in fact, CsI is hygroscopic and water contamination degrades its QE. Secondly, degradation of QE is also caused by intense bombardment (0.2 $mC/cm^{2}$) of ions generated by the multiplication process in gaseous detectors \cite{Hoedlmoser}. The quest for an alternative material based PC free from these limitations has therefore been a prime goal for the R\&D of the future EIC RICH. In this article, we report about the preliminary studies performed with THGEM detectors coated with novel PC material, based on hydrogenated nanodiamond powder. 
\section{Nanodiamond particles as an alternative to CsI}
Low electron affinity ($0.1~eV$) and wide band gap ($6.2~eV$) makes CsI the mostly used material in the field of UV PCs ~\cite{CsI}. Diamond has a band gap of $5.5~eV$ and low electron affinity of $0.35$-$0.50~eV$ and it also exhibits chemical inertness, radiation hardness and good thermal conductivity~\cite{NDRep-1}. Additionally, hydrogenation of diamond surface lowers the electron affinity down to -$1.27~eV$. The negative electron affinity favours an efficient escape into vacuum of the generated photoelectrons 
without an energy barrier at the surface~\cite{NDRep-1}. Recently, a novel procedure has been developed ~\cite{NDRep-1, NDreport} in the Thin  Film  Lab  (Dipartimento  di  Fisica-University of Bari) and MWPECVD (The Microwave Plasma Enhanced Chemical Vapor Deposition ) lab (CNR-NANOTEC of Bari), where hydrogenation of nanodiamond particles is performed prior to the coating and strikingly high and stable QE values have been reported. A comparison between the quantum efficiency of CsI and nanodiamond can be made from literature~\cite{CsI,NDRep-1}.  
\section{Pre-characterization and coating}
\subsection{Characterization before Coating}
\par As starting exercise of our R\&D studies, we coated four THGEM detectors. THGEMs are electron multipliers of the MPGD technology. They are derived from GEM architecture with thicker dielectric material, a PCB in our case, in between two electrode layers.
\begin{figure}
\begin{minipage}[c]{0.6\textwidth}
    \includegraphics[width=0.8\textwidth]{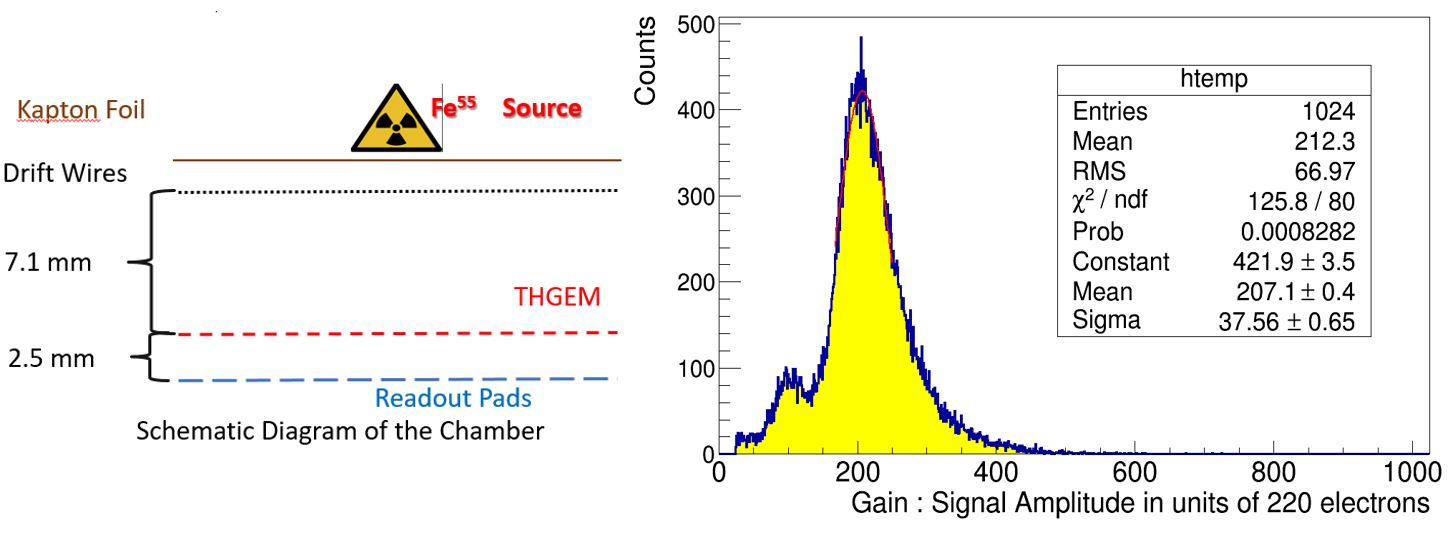}
\end{minipage}\hfill
\begin{minipage}[c]{0.48\textwidth}
        \caption{(left) The schematic of our detector set up. (right)
        A typical ${}^{55}Fe$ spectrum obtained in  $Ar:CO_{2}~=~70:30$ gas mixture, where the peak position provides the effective gain of the detector, in unit of ADC channels. Each ADC channel corresponds to 220 electrons, which is approximately the number of electrons produced in Ar by a fully converted X-ray from ${}^{55}Fe$ source.}
       \label{fig:spectra}
    \end{minipage}
\end{figure}
We have fully characterized the THGEMs in the Trieste INFN laboratory before performing the coating procedure at the Thin Film Lab (Dipartimento di Fisica-University of Bari) and MWPECVD Lab (CNR-NANOTEC of Bari). The pre-characterization ensures a well defined reference to compare performance before and after the coating. The THGEMs have $30\times30~mm^{2}$ active area and the Cu surface is Ni-Au coated. 
The hole diameter is 0.4~mm and the pitch is 0.8~mm. The thickness is 470~$\mu$m including 35~$\mu$m of Cu on both faces. THGEMs have different rim size, where the rim is the clearance 
ring around  the hole edge: $\le5\mu$m (no rim), $\sim10~\mu$m and $\sim20~\mu$m.
\par 
Figure (\ref{fig:spectra}) displays the schematic set-up and a typical spectrum 
of the  pre-characterization procedure. The peak position of the spectrum provides the effective gain of the detector. Conventionally, drift field is defined as the applied field between the drift wires and the top layer of the THGEM. Similarly, the induction field is the applied field in between the bottom layer of THGEM and the readout pad. During characterization we have determined and applied optimized drift and induction fields, to study the effective gain of the detector applying different voltage across the THGEM. This characterization procedure had been repeated for each of the THGEMs before coating. After the coating, we have studied the performance of the THGEMs in the same voltage configuration to appreciate the variations due to the coating.
\subsection{Coating procedure}
\begin{figure}
\begin{minipage}[c]{0.6\textwidth}
    \includegraphics[width=0.8\textwidth]{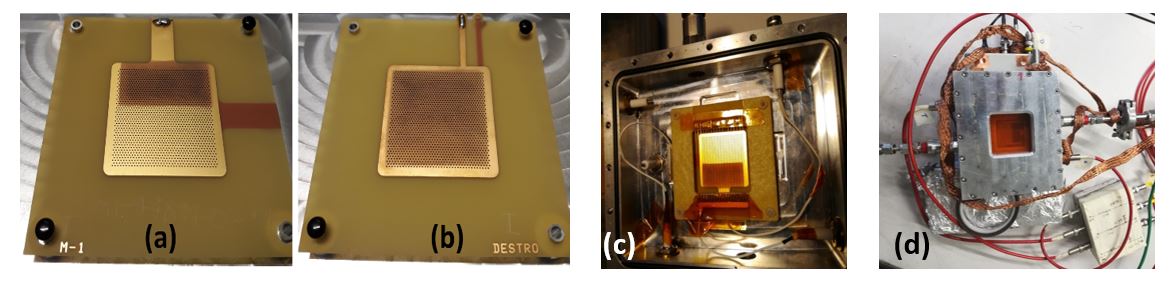}
\end{minipage}\hfill
\begin{minipage}[c]{0.48\textwidth}
        \caption{ THGEMs with nanodiamond particles covering (a) half and (b)
        fully the surface area; (c) coated THGEM installed in the chamber;
        (d) the test chamber with gas flow. 
}
        \label{fig:coating}
    \end{minipage}
\end{figure}

MWPECVD diamonds are used for thermo-ionic current generation and UV PCs because they exhibit better stability than CsI upon exposure to air ~\cite{coating-II, CsI_QE_deg}. It is also expected to be stable under irradiation. The standard diamond deposition is performed at high temperature, namely $800^{O}$C at least and only small areas can be covered. Furthermore, the high temperature prevents the use of materials intolerant to this temperature. In order to get rid of these limitations, a novel and low-cost technique has been developed in the Thin  Film  Lab  (Dipartimento  di  Fisica-University of Bari) and MWPECVD Lab (CNR-NANOTEC of Bari), where nanodiamond particles with an average grain size of 250 nm supplied by Diamonds\&Tools srl are used for the coating. The hydrogenated nanodiamond is obtained by treating the as-received powder in $H_{2}$ microwave plasma for 1h. The nanodiamond and hydrogenated nanodiamond particles were separately dispersed in deionized water, sonicated for 30 min by a Bandelin Sonoplus HD2070 system and then sprayed  at $120^{0}$C on the THGEMs by coating half or fully their surface area (Fig.~{\ref{fig:coating}}) with 400 or 800 spray pulses, respectively, giving a granular layer formed by nanodiamond particles. Further details can be found elsewhere~\cite{coating-I, coating-II}.  It is worth to mention that extending the spray technique to larger surfaces is relatively easy. 
\section{Post-coating Characterization}
\begin{figure}
\begin{minipage}[c]{0.5\textwidth}
    \includegraphics[width=\textwidth]{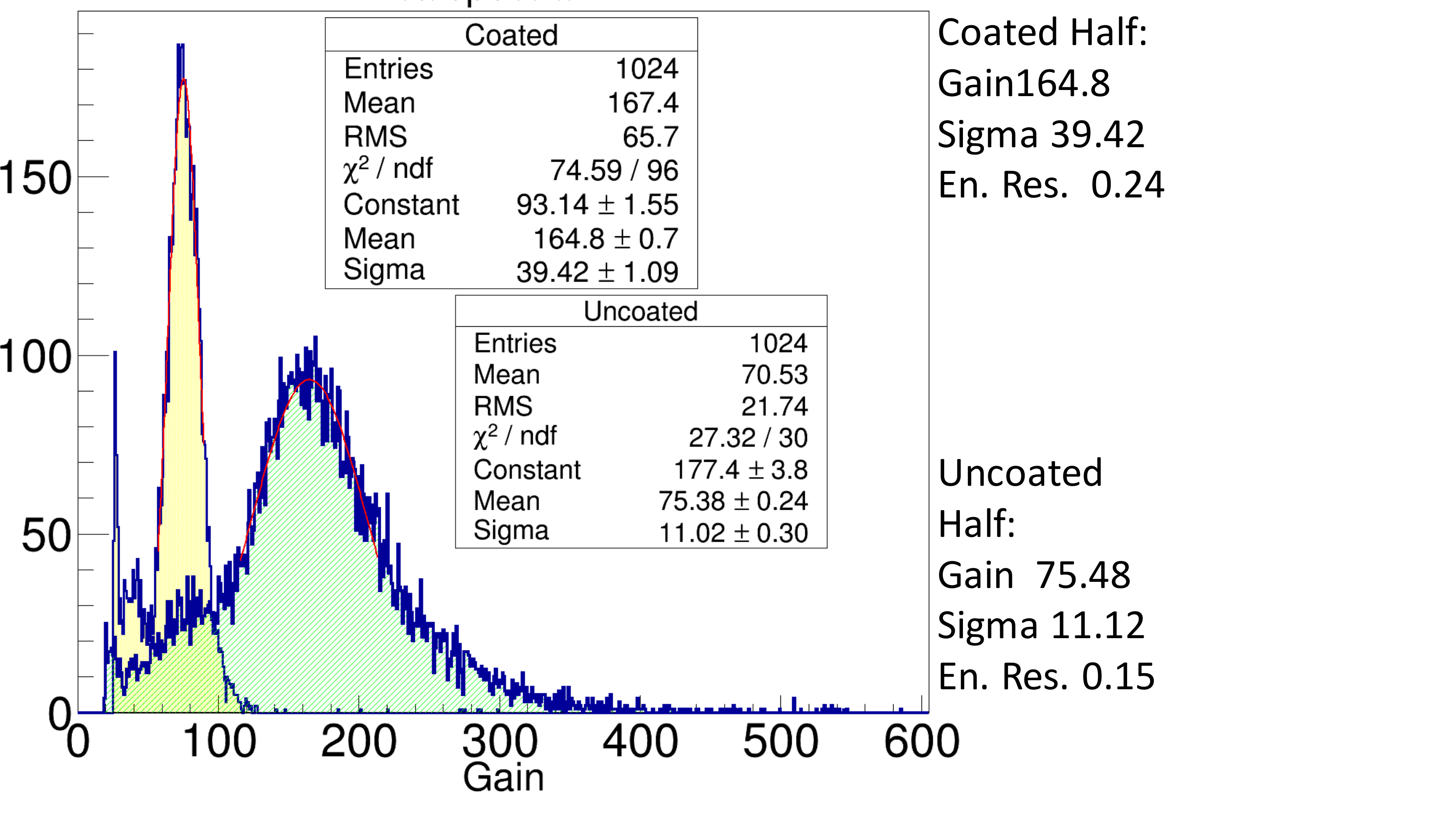}
\end{minipage}\hfill
\begin{minipage}[c]{0.48\textwidth}
        \caption{ Gain behavior of THGEM with 20~$\mu$m rim, half-coated with nanodiamond. It is clearly shown that the gain in the coated part is almost two times higher than that in the the uncoated part.}
        \label{fig:gain-increment}
    \end{minipage}
\end{figure}

During the post coating characterization, we have observed several interesting
effects. The gain in the coated part is higher than the uncoated part. The
factor of increment is however different on THGEMs with different rim size. 
The THGEMs with rim size $\sim20~\mu$m coated with non-hydrogenated nanodiamond
showed a $\sim$2.5~times increment in the gain.
\begin{figure}
\begin{minipage}[c]{0.5\textwidth}
    \includegraphics[width=\textwidth]{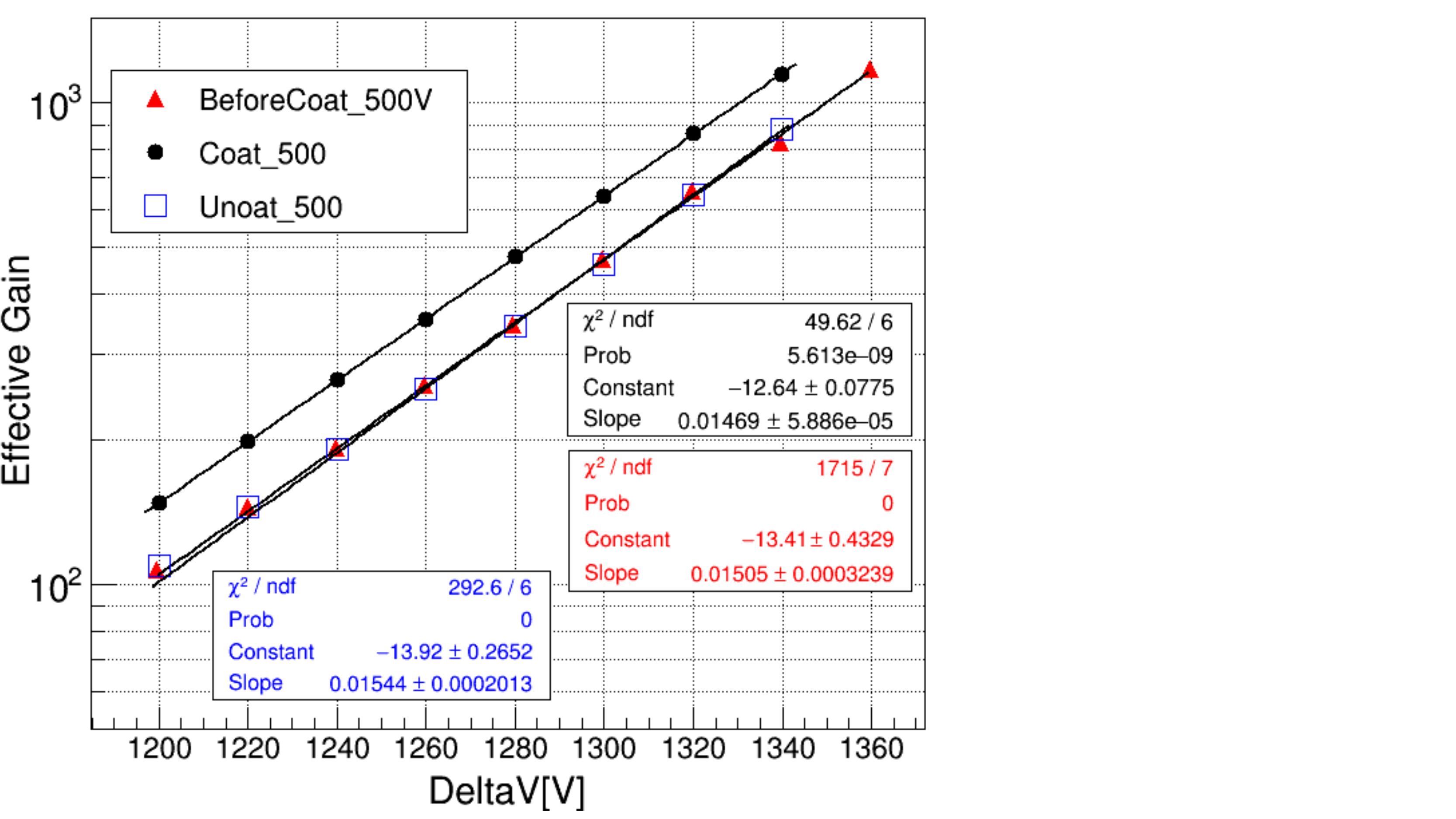}
\end{minipage}\hfill
\begin{minipage}[c]{0.48\textwidth}
        \caption{ Gain versus applied voltage across a THGEM with no rim, half-coated with nanodiamond. The gain measured in the coated part
        is compared with that measured before coating and after coating, in the uncoated portion of the THGEM.}
        \label{fig:gainwithvolt}
    \end{minipage}
\end{figure}
In case of a THGEM with no rim, the rise in the gain is only a factor $\sim$1.3. Figures~{\ref{fig:gainwithvolt} and \ref{fig:gain-increment}} show the increment in the gain. These gain variations are potentially related to a different charging up of the dielectric surface when the coating is present, an interesting mechanism to be confirmed by further studies.
\par
\begin{figure}
\begin{minipage}[c]{0.5\textwidth}
    \includegraphics[width=\textwidth]{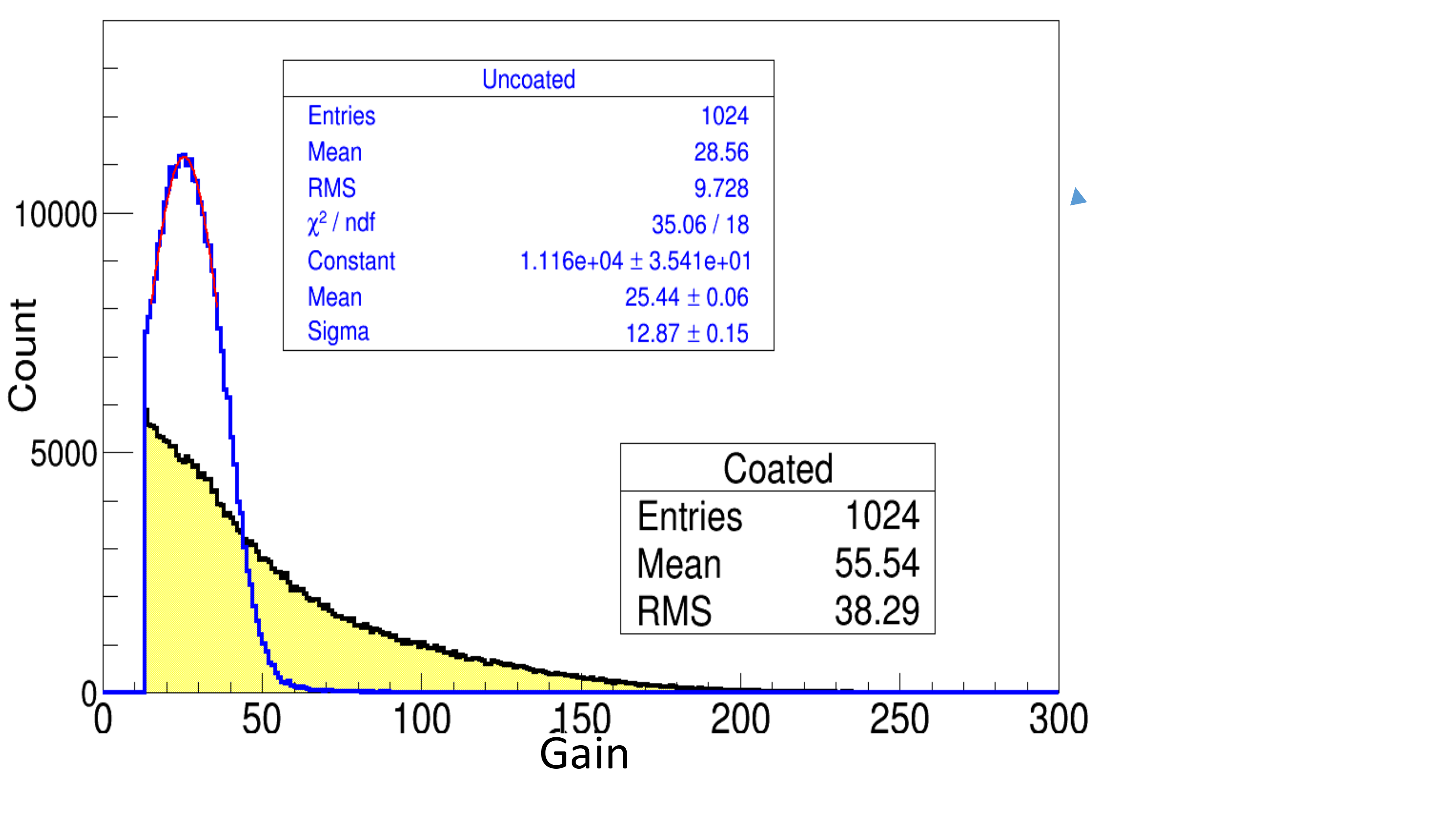}
\end{minipage}\hfill
\begin{minipage}[c]{0.48\textwidth}
        \caption{Gain comparison of hydrogenated nanodiamond coated and uncoated part of a no rim THGEM.}
        \label{fig:gain-HND}
    \end{minipage}
\end{figure}

 
The hydrogenated  nanodiamonds coated THGEMs exhibit lower electrical stability after the
coating. In fact, we have observed discharges for the fully coated hydrogenated nanodiamond THGEMs at 
the voltages where the same THGEMs before coating coud be operated in a spark free regime.
The estimated gain for the half-coated  hydrogenated nanodiamond THGEMs and no rim,
shows a increment of factor $\sim2$ and a lower electrical stability (Fig.~\ref{fig:gain-HND}).
These observations point to a potential resistivity issue related to the 
hydrogenated nanodiamond coating, to be confirmed by further studies.
\section{Conclusion}
A detailed preliminary characterization of THGEMs before and after coating with 
nanodiamond particles has been performed. We have observed two interesting phenomena by comparing the respective performance results: a) lower electrical stability of the THGEMs coated by hydrogenated nanodiamond particles w.r.t. the non-coated ones;  b) larger gain in the coated part for both the hydrogenated and non-hydrogenated films. A systematic study is needed to explain these interesting preliminary observations and to improve our knowledge of the effect of the nanodiamond deposition on THGEMs substrates and of the photo conversion mechanism. Once fully understood the technology may be an appealing solution where VUV photoconversion is needed and  may 
have interesting  applications  in  experimental  particle 
physics as well as in many other scientific and industrial fields.  
\section{Acknowledgement}
This work is partially supported by the H2020 project AIDA-2020, GA no. 654168.


\section{References}
\begin{thebibliography}{}
\bibitem{EIC}
Electron Ion Collider: The Next QCD Frontier; Eur.Phys.J. A52 (2016) no.9, 268; arXiv:1212.1701
\bibitem{windowless-RICH}
M. Blatnik et al., Performance of a Quintuple-GEM Based RICH Detector Prototype, IEEE NS 62 (2015) 3256.
\bibitem{HBD}
W.Anderson et al., Nucl. Instr. and Meth. A 646 (2011), 35.
\bibitem{THGEM-RICH2018}
F.Tessarotto et al., The Hybrid MPGD-based photon detectors of COMPASS RICH-1, talk at RICH2018, Moscow, Russia, 29 July - 4 August 2018
\bibitem{Hoedlmoser}
H. Hoedlmoser et al., Long term performance and ageing of CsI photocathodes for the ALICE/HMPID detector, NIMA 574 (2007) 28
\bibitem{CsI}
F.Piuz, Ring Imaging CHerenkov systems based on gaseous photo-detectors: trends and limits around particle accelerators, NIMA 502 (2003) 76
\bibitem{NDRep-1}
L. Velardi, A. Valentini, G. Cicala, Highly efficient and stable UV photocathode based on nanodiamond particles, Appl. Phys. Lett. 108 (2016) 083503.
\bibitem{coating-I}
G. Cicala, A. Massaro, L. Velardi, G. S. Senesi, A. Valentini, Self-Assembled Pillar-Like Structures in Nanodiamond Layers by Pulsed Spray Technique, ACS Appl. Mater. Interfaces 6 (2014) 21101-21109.
\bibitem{coating-II}
L. Velardi, A. Valentini, G. Cicala, UV photocathodes based on nanodiamond particles: effect of carbon hybridization on the efficiency, Diam. Relat. Mater. 76 (2017) 1-8.
\bibitem{CsI_QE_deg}
A.S. Tremsin, S. Ruvimov, O.H.W. Siegmund, Structural transformation of CsI thin film photocathodes under exposure to air and UV irradiation, Nucl. Inst. Methods Phys. Res. A 447 (2000) 614.
\bibitem{NDreport}
A. Valentini, D. Melisi, G. De Pascali, G. Cicala, L. Velardi, A. Massaro, “High-efficiency nanodiamond-based ultraviolet photocathodes”, 30-03-2017 Patent n. WO 2017/051318 A9; International Patent n. PCT/IB2016/055616 of September 21, 2016; National Patent Italia - n. 102015000053374 del 21 Settembre 2015, Istituto Nazionale di Fisica Nucleare e Consiglio Nazionale delle Ricerche.
\end{thebibliography}
\end{document}